# Primordial Black Holes in the Solar System

Valentin Thoss[1,2,3] and Andreas Burkert[1,2,3]
[1] Universitäts-Sternwarte, Ludwig-Maximilians-Universität München, Scheinerstr. 1, 81679 Munich, Germany; vthoss@mpe.mpg.de
[2] Max-Planck-Institut für Extraterrestrische Physik, Gießenbachstraße 1, 85748 Garching, Germany
[3] Excellence Cluster ORIGINS, Boltzmannstraße 2, 85748 Garching, Germany

## Abstract

If primordial black holes (PBHs) of asteroidal mass make up the entire dark matter, they could be detectable through their gravitational influence in the solar system. In this work, we study the perturbations that PBHs induce on the orbits of planets. Detailed numerical simulations of the solar system, embedded in a halo of PBHs, are performed. We find that the gravitational effect of the PBHs is dominated by the closest encounter. Using the Earth–Mars distance as an observational probe, we show that the perturbations are smaller than the current measurement uncertainties and thus PBHs are not directly constrained by solar system ephemerides. We estimate that an improvement in the ranging accuracy by an order of magnitude or the extraction of signals well below the noise level is required to detect the gravitational influence of PBHs in the solar system in the foreseeable future.

*Unified Astronomy Thesaurus concepts:* Solar system evolution (2293); Solar system (1528); Primordial black holes (1292); Milky Way dark matter halo (1049); Dark matter (353); Close encounters (255); Orbital anomalies (1176); Gravitation (661); Black holes (162)

## 1. Introduction

Primordial black holes (PBHs) as a dark matter (DM) candidate have been studied for half a century (B. J. Carr & A. M. Green 2024), gaining special attention after the first direct detection of gravitational waves from binary black hole mergers (B. P. Abbott et al. 2016). By now, a large number of constraints have been derived that limit the fraction of DM that can be in the form of PBHs (B. Carr et al. 2021). However, these constraints come with some uncertainty and can even disappear entirely, as was recently demonstrated for light PBHs (A. Alexandre et al. 2024; V. Thoss et al. 2024).

The asteroid-mass window ($M_{\rm PBH} \in [10^{17}, 10^{23}]$ g) has been studied with particular interest, as it remains a viable parameter region for PBHs. Within this mass range, their interaction with stars, neutron stars, and white dwarfs has been studied as a pathway to detect them or constrain their DM fraction. However, many of the bounds that were obtained in this way are disputed for various reasons (see B. Carr et al. 2021 for an overview).

Another approach to studying PBHs of asteroidal mass is through their effect within the solar system. It has been suggested to look for craters as a signature of PBH collisions with moons and planets (A. Yalinewich & M. E. Caplan 2021; M. E. Caplan et al. 2023). Other work focuses on the gravitational effects of PBHs. This includes perturbations to the orbits of moons and planets (Y.-L. Li et al. 2023; T. X. Tran et al. 2024), satellite constellations (B. Bertrand et al. 2023), and future space-based gravitational-wave detectors such as LISA (A. W. Adams & J. S. Bloom 2004; N. Seto & A. Cooray 2004). So far these are mostly proof-of-concept studies that suggest that an accurate model for solar system ephemerides, combined with a sophisticated data analysis, will make the detection of individual asteroid-mass PBHs feasible.

T. X. Tran et al. (2024) showed that if the extraction of signals with an amplitude of $10^{-4}$ relative to the noise is achieved, then there will be a significant rate of detectable PBH encounters for $10^{18}\,{\rm g} < M_{\rm PBH} < 10^{23}\,{\rm g}$ with the current observational accuracy.

Recently, A. Loeb (2024) argued that PBHs within most of the asteroid-mass window are already excluded from making up the entire DM, based on their perturbations of solar system bodies. The result was obtained by considering the Poissonian fluctuation of the number of PBHs within a given radius $R$ from the Sun. The rate of change of the total PBH mass, enclosed within $R$, was compared to an observational constraint on the rate of change of the solar mass. This approach assumes that the total mass of PBHs within $R$ has a gravitational effect similar to a point mass and therefore can be added directly to the mass of the Sun. From this, A. Loeb (2024) concluded that, for $R = 50$ au, PBHs cannot make up the entire DM in the mass range $M_{\rm PBH} \in [6 \times 10^{18}, 10^{22}]$ g. However, J. M. Cline (2024) has noted that the choice of $R = 50$ au is not justified and that one cannot easily rule out PBHs as a DM candidate.

Because of the far-reaching consequences of the results obtained by A. Loeb (2024), it is necessary to investigate it in more detail. The key question is whether PBHs can induce detectable perturbations on the orbits of solar system objects (SSOs). In this work we present results from $N$-body simulations of the solar system, embedded in a halo of PBHs. We study the orbital perturbations that these compact objects induce on planets in the solar system. Our methods are presented in Section 2, and our results are shown in Section 3. In Section 4 we provide a discussion of our results and show that the perturbations are dominated by the closest encounter rather than the Poissonian fluctuations in the PBH number density studied by A. Loeb (2024). We conclude with a summary in Section 5.

## 2. Methods

Our goal is to simulate the perturbations induced on the orbits of SSOs by a halo of asteroid-mass PBHs. In this section

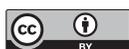

**Table 1**
Simulation Parameters, Showing the Adopted PBH Mass and Number, the Size of the Simulation Box, the Total Number of Runs, and the Simulation Time

| $M_{PBH}$ (g) | $N_{PBH}$ | $R_{box}$ (au) | $N_{runs}$ | $t$ (yr) |
|---|---|---|---|---|
| $10^{18}$ | 18525 | 200 | 1000 | 1 |
| $10^{19}$ | 6252 | 300 | 1000 | 1 |
| $10^{20}$ | 1482 | 400 | 3000 | 1 |
| $10^{21}$ | 500 | 600 | 9000 | 1 |
| $10^{20}$ | 2110 | 450 | 500 | 20 |

we briefly describe our simulation methods and relevant quantities.

We use a second-order Leapfrog integrator with a fixed time step to simulate the motions of the Sun, the eight planets, and Earth's Moon. Other SSOs are neglected, as we are only interested in the relative perturbation of planets in the solar system. The goal of this work is not to make accurate predictions on the absolute positions of the bodies of the solar system; rather, we are interested in the relative perturbation of their position $\delta r(t)$. We do not expect the smaller bodies to have a sizable effect on these perturbations. For the same reason, we do not treat finite-size effects or relativistic corrections, as these will only affect the perturbations at second order. We refer to T. X. Tran et al. (2024) for a more detailed discussion on these effects.

The initial conditions for the SSOs are obtained from the Horizons System by JPL, based on the DE441 model (R. S. Park et al. 2021). The solar system is embedded in a halo of PBHs, homogeneously filling a box of size $L_{box}$ with periodic boundary conditions. The halo is populated with $N_{PBH} = L_{box}^3 \rho_{CDM}/M_{PBH}$ PBHs of random positions, thus assuming that PBHs make up the entire DM with a monochromatic mass distribution. During the simulations, the gravitational forces are only computed for PBHs within a sphere of radius $R_{box} = L_{box}/2$. This enables the comparison with the case of a smooth spherical particle DM, where the potential is analytically tractable. The PBHs are given a Maxwellian velocity distribution with a dispersion $\sigma_v = 185$ km s$^{-1}$, as well as an additional component due to the rotation of the galactic disk. The latter has a magnitude of $v_\odot = 230$ km s$^{-1}$, an angle of $60°$ with respect to the ecliptic plane and a direction such that the relative motion between the PBHs and Earth is maximal on June 1 (K. Freese et al. 2013). These parameters lead to a relative velocity of the PBHs of $v_{rms} \approx 279$ km s$^{-1} \approx 59$ au yr$^{-1}$. Here $\rho_{CDM} = 7 \times 10^{-25}$ g cm$^{-3}$ is assumed to facilitate comparisons with A. Loeb (2024).

To investigate the parameter space of the asteroid-mass window, we choose $M_{PBH} \in [10^{18}, 10^{19}, 10^{20}, 10^{21}]$ g. Note that we discuss how our results can be extrapolated to other PBH masses. For each value of the mass $M_{PBH}$ a large number of simulation runs ($\mathcal{O}(1000)$; see Table 1) are performed, each over a physical time span of 1 yr, to account for the randomness of the encounters with the solar system bodies. These simulations help us to understand the dependence of the perturbation strength on the black hole mass $M_{PBH}$. To answer the question whether PBHs can cause detectable perturbations in the solar system, 500 simulations with $M_{PBH} = 10^{20}$ g are carried out over a time of 20 yr, which is roughly the time span for which the most precise ranging data in the solar system have been available. We checked for each value of $M_{PBH}$ that both the numerical time step and the value of $R_{box}$ do not significantly affect our results. To reduce computational cost, the gravitational force is only calculated for the solar system bodies, whereas the PBHs move on straight trajectories. This is a reasonable approximation owing to the high velocity of the PBHs. We nevertheless performed additional simulations including the gravitational force for the PBHs and found that it only changes our results below the percent level. The exact parameters of our simulation ensemble can be found in Table 1.

The main quantity of interest is the perturbation that the PBHs induce on the distance between Earth and a given SSO,

$$\frac{\delta r}{r}(t) = \frac{|\tilde{\mathbf{r}}_{earth}(t) - \tilde{\mathbf{r}}_{SSO}(t)| - |\mathbf{r}_{earth}(t) - \mathbf{r}_{SSO}(t)|}{|\mathbf{r}_{earth}(t) - \mathbf{r}_{SSO}(t)|}, \quad (1)$$

where $\tilde{\mathbf{r}}$ indicates the position in a simulation with PBHs and $\mathbf{r}$ refers to the comparative simulation with a smooth DM halo. Naively, one might assume that the comparative simulation setup must be a solar system without DM. However, due to our numerical setup with a spherical PBH halo, centered around the barycenter of the solar system, there is a small additional acceleration for each SSO owing to the presence of DM,

$$\ddot{\mathbf{r}} = -\frac{4\pi G \rho_{CDM}}{3}\mathbf{r}. \quad (2)$$

Adding this term in the comparative simulation is necessary, as we are interested in the perturbations induced by the PBHs and not in the numerical effect of the additional mass within the solar system. We want to mention that in this work we will focus on the perturbation of the magnitude of the vector,

$$\frac{|\delta \mathbf{r}|}{r}(t) = \frac{|(\tilde{\mathbf{r}}_{earth}(t) - \tilde{\mathbf{r}}_{SSO}(t)) - (\mathbf{r}_{earth}(t) - \mathbf{r}_{SSO}(t))|}{|\mathbf{r}_{earth}(t) - \mathbf{r}_{SSO}(t)|}, \quad (3)$$

which has the advantage of being strictly positive, whereas $\delta r/r$ typically oscillates between $-|\delta r|/r$ and $|\delta r|/r$ within one synodic orbital period of the SSO considered. While the quantity $|\delta r|/r$ is more difficult to observe, it provides a reasonable estimate of the largest observable perturbation $\delta r/r$ per orbital period, as we show in Section 3.

To study the possibility of detecting or constraining PBHs, we compare our simulation results to observational data. We can relate the magnitude of the induced perturbations to the measurement accuracy for solar system bodies. Currently available data for various solar system bodies allow the detection of perturbations as small as $\delta r/r \sim 10^{-11}$. At the moment, the most precise data are obtained for the Moon ($\mathcal{O}(1$ mm$)$), by lunar laser ranging (J. B. R. Battat et al. 2023; N. R. Colmenares et al. 2023), and for Mercury ($\mathcal{O}(0.7$ m$)$) and Mars ($\mathcal{O}(0.7$ m$)$) through various orbiters (R. S. Park et al. 2021). While the SSOs have been monitored for a long time, submeter accuracy for Mars and Mercury has only been achieved in the past two decades. In this work we will focus on the distance between Earth and Mars, as it has been measured with high accuracy and is least susceptible to effects not considered in this work. These include, most notably, finite-size effects for the Moon and relativistic point-mass effects for Mercury (see T. X. Tran et al. 2024 for some estimates).

Finally, let us emphasize that our work does not aim to describe singular encounters with PBHs (for which we refer to B. Bertrand et al. 2023; Y.-L. Li et al. 2023; T. X. Tran et al. 2024) but instead

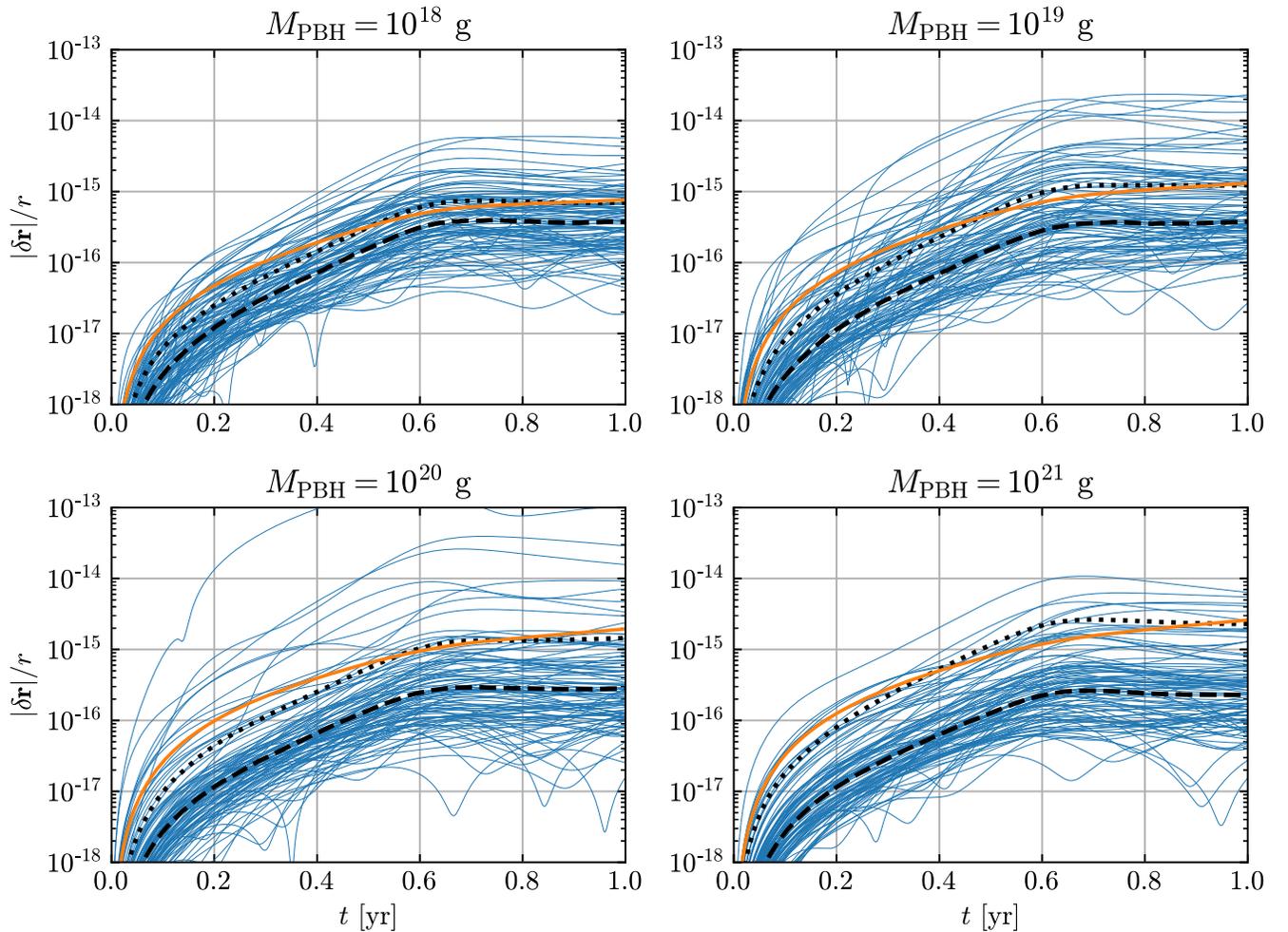

**Figure 1.** Perturbation of the vector between Earth and Mars, induced by PBHs for different values of their mass $M_{\rm PBH}$. Each blue line corresponds to one simulation run. We only show 100 simulation runs in each case to make comparisons easier and improve readability. The black dashed/dotted line indicates the median/mean value obtained from all simulations. The analytical prediction of the impulse model, given by Equation (8) for $b_{\rm max} = 3b_1$, is shown as an orange line.

focuses on the cumulative effect of a whole halo of PBHs on the dynamics of the solar system.

## 3. Results

Figure 1 shows the perturbations that are induced on the vector between Earth and Mars over a time span of $t = 1$ yr for $M_{\rm PBH} \in [10^{18}, 10^{19}, 10^{20}, 10^{21}]$ g. Due to the random initial conditions of the PBHs for each run, there is a significant variance in the value of $|\delta \mathbf{r}|/r$. The median value of the perturbations induced by a halo of PBHs grows over time and does not show a significant dependence on the black hole mass $M_{\rm PBH}$. The mean value of $|\delta \mathbf{r}|/r$ is higher than the median and increases slowly with $M_{\rm PBH}$. The growth of perturbations can be understood using the analytical model derived in Section 4.1 and refined in Section 4.3, which is based on the impulse approximation. The predicted value for the mean perturbation strength is shown in Figure 1 and agrees well with the simulation results.

In general, the model predicts a logarithmic dependence on the black hole mass when the distance $b$ to the nearest PBH is much larger than the distance $r$ to the respective SSO. In contrast, in the limit $b \ll r$ we expect a power-law dependence $|\delta \mathbf{r}|/r \sim \sqrt{M_{\rm PBH}}$. To test this, we show the mean perturbation of the vector between Earth and each planet at $t = 1$ yr in the left panel of Figure 2, together with the prediction from the analytical model. Indeed, the perturbation strength depends more strongly on $M_{\rm PBH}$ in the limit $b \ll r$. Overall, the predictions from the analytical model are within a factor of two of the simulation results. It is noticeable that the agreement is better for the inner planets compared to the outer ones. We note that the mean perturbation strength that we obtain from the simulations is subject to some statistical uncertainty because of the finite number of simulations that have been performed. We discuss this in more detail in Section 4.2.

For a subset of all simulations performed we also evaluate the distance of each SSO to the closest PBH. In the right panel of Figure 2 the perturbation strength $|\delta \mathbf{r}|/r$ at $t = 1$ yr is plotted against the smallest distance $b_{\rm min}$ of Mars and Earth to a PBH within the simulation run. The dashed lines correspond to an analytical estimate of the perturbation by a single encounter with an impact parameter $b_{\rm min}$, given by Equation (7). The good agreement demonstrates that the perturbation strength is mostly determined by the closest encounter. Note that for $M_{\rm PBH} = 10^{18}$ g the typical distance to the nearest PBH is on the order of $b \sim 1$ au.

A statistical analysis of the perturbations is presented in Figure 3, where we show the probability distribution $p(|\delta \mathbf{r}|/r)$ of the perturbations of the Earth–Mars vector at $t = 1$ yr in the left panel. The functions are obtained by a Gaussian kernel

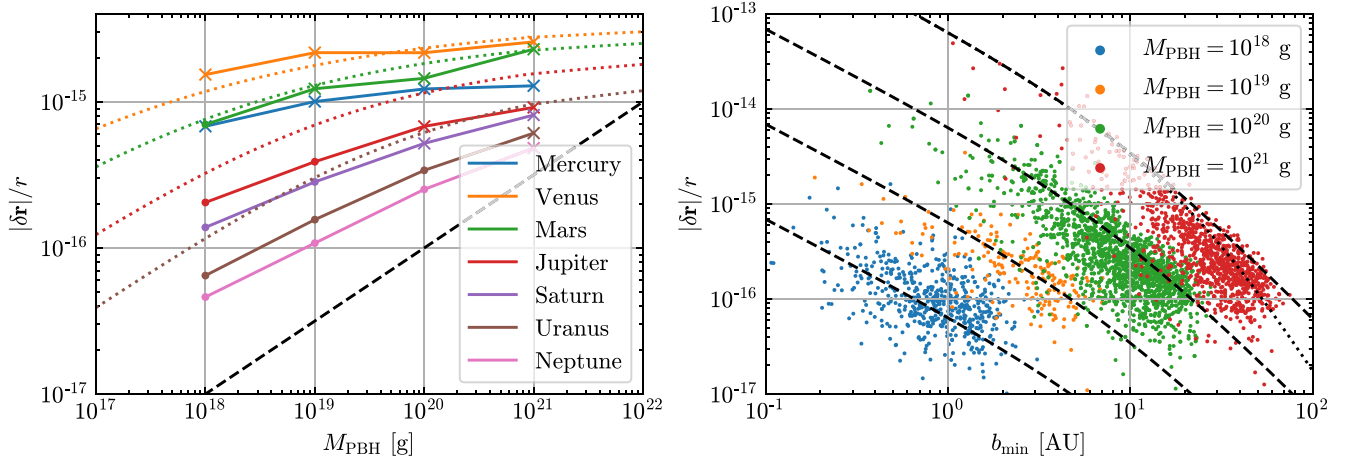

**Figure 2.** Left: mean perturbation strength $|\delta \mathbf{r}|/r$ at $t = 1$ yr as a function of $M_{\rm PBH}$ for the vector between Earth and each planet. The crosses indicate that the distance $b$ to the closest PBH is larger than the distance $r$ between Earth and the respective planet in that case, whereas the points indicate that $b < r$. Colored dotted lines show the analytical prediction from Equation (17) for selected planets. The black dashed line indicates a power-law slope of 0.5. Right: perturbation strength $|\delta \mathbf{r}|/r$ of the Earth–Mars vector, evaluated at $t = 1.0$ yr, vs. the minimum impact parameter $b_{\rm min}$ observed within the simulation period. The black dashed lines correspond to Equation (7). The dotted line for $M_{\rm PBH} = 10^{21}$ g takes into account the finite distance traveled by the PBHs within the simulation time and is given by Equation (14).

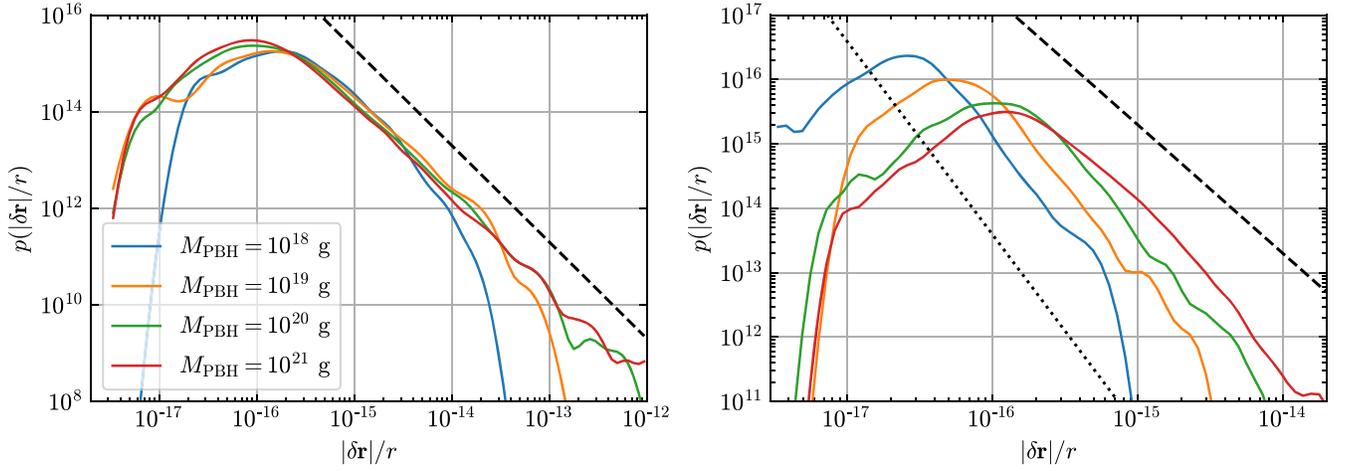

**Figure 3.** Left: probability distribution function of the perturbation $|\delta \mathbf{r}|/r$ of the Earth–Mars vector, evaluated at $t = 1.0$ yr for different values of the PBH mass $M_{\rm PBH}$. The black dashed line indicates a power law with slope of $-2$. Right: probability distribution function of the perturbation $|\delta \mathbf{r}|/r$ of the Earth–Neptune vector, evaluated at $t = 1.0$ yr for different values of the PBH mass $M_{\rm PBH}$. The black dashed and dotted lines indicate a power law with slope of $-2$ and $-3$, respectively.

density estimate. Interestingly, there is no strong dependence on the PBH mass, and for $M_{\rm PBH} \geqslant 10^{19}$ g we observe a power-law tail with a slope of $-2$. As we show in Section 4.2, a power-law slope of $-2$ is expected if the distance to the nearest PBH $b$ is much larger than the distance $r$ to the respective SSO, here the Earth–Mars distance. This is the case for $M_{\rm PBH} \geqslant 10^{19}$ g (see the right panel of Figure 2). In contrast, if $b \ll r$, in which case there will be many PBHs between Earth and the respective SSO, then our model predicts a slope of $-3$. To study this limit, we show the probability distribution $p(|\delta \mathbf{r}|/r)$ of the perturbations of the Earth–Neptune vector at $t = 1$ yr in the right panel of Figure 3. As the distance between Earth and Neptune is much larger compared to Earth–Mars, a slope of $-3$ is expected for $M_{\rm PBH} \leqslant 10^{20}$ g (see the right panel of Figure 2), which is indeed what we find. Note that we observe a stronger mass dependence in that case, with an increase in perturbation strength $|\delta \mathbf{r}|/r$ toward larger values of $M_{\rm PBH}$. This is consistent with our analytical model, which predicts $|\delta \mathbf{r}|/r \sim \sqrt{M_{\rm PBH}}$ in the limit $b \ll r$.

Finally, Figure 4 presents the results from 500 simulations performed over a longer time span of 20 yr for $M_{\rm PBH} = 10^{20}$ g.

In the left panel we display the perturbation $|\delta \mathbf{r}|/r$ of the vector between Earth and Mars. A notable difference to Figure 1 is that the perturbations oscillate with a period of roughly 2 yr, which corresponds to the synodic orbital period of Mars. We add an estimate for a $3\sigma$ detection limit by assuming that $\sigma_r \approx 70$ cm for the Earth–Mars distance (R. S. Park et al. 2021). Out of all 500 simulation runs, only 4 exceed this limit at least once within 20 yr, implying a ~1% chance of detection with a confidence level of $3\sigma$.

In the right panel of Figure 4 we show the mean perturbation strength $|\delta \mathbf{r}|/r$ of all simulation runs for the vector between Earth and each planet. Notably, $|\delta \mathbf{r}|/r$ is very similar for the inner planets, while it decreases for the outer planets as the distance to Earth grows. Since the smallest typical impact parameter for $M_{\rm PBH} = 10^{20}$ g and $t = 20$ yr is $b_1 \approx 3.4$ au (see Equation (9)), we are in the limit $b > r$ for the inner planets and $b < r$ for the outer planets. The analytical model derived in Section 4.1 predicts a constant value of $|\delta \mathbf{r}|/r$ in the case $b > r$ and $|\delta \mathbf{r}|/r \sim 1/r$ otherwise. Therefore, the lower perturbation strength for the outer planets is expected.

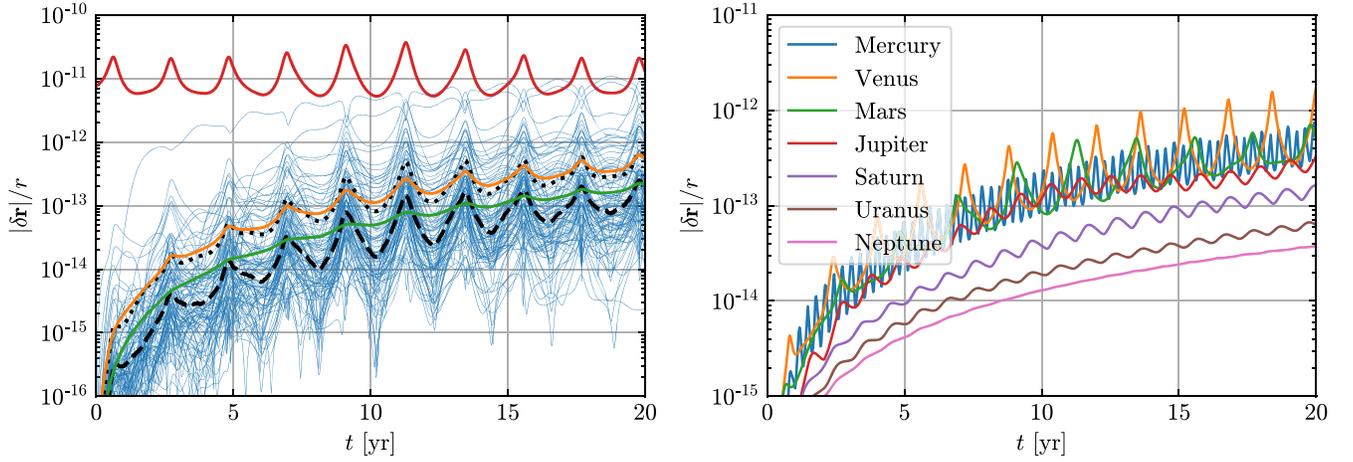

**Figure 4.** Left: perturbation of the vector between Earth and Mars induced by PBHs for $M_{\rm PBH} = 10^{20}$ g as a function of time. Each blue line corresponds to one simulation run. We only show 100 simulation runs to improve readability. The black dashed/dotted line indicates the median/mean value obtained from all 500 simulations. The analytical prediction of the impulse model, given by Equation (8) for $b_{\rm max} = 3b_1$, is shown as an orange line. The green line shows a model for the median perturbation strength, given by Equation (19). The red line is an estimate of the $3\sigma$ observational detection limit. Right: mean perturbation strength $|\delta r|/r$ as a function of time for the vector between Earth and each planet.

In the results presented in this study, we analyze the perturbation of the vector $|\delta \boldsymbol{r}|$ between Earth and other SSOs. However, only the perturbation of the distance $\delta r$ can be observed with high accuracy. We argue that this should not affect our results, as $\delta r$ typically oscillates between $-|\delta \boldsymbol{r}|$ and $|\delta \boldsymbol{r}|$, and thus our results can be regarded as an upper bound of the observable perturbation since $|\delta r| \leqslant |\delta \boldsymbol{r}|$ by definition. To test this statement, we compare the values of $\delta r$ and $\delta \boldsymbol{r}$ for the Earth–Mars pair. The data from the last synodic orbital period from the 500 simulations performed over a time span of $t = 20$ yr are being used. We find that 87% of the simulations reach $|\delta r| > |\delta \boldsymbol{r}|/2$ at least once and 72% of the simulations reach $|\delta r| > |\delta \boldsymbol{r}|/2$ during at least 75% of the respective time span. In addition, 50% of the simulations reach $|\delta r| > 0.9|\delta \boldsymbol{r}|$ at least once within the orbital period. This justifies our approach of studying the perturbation of the vector $\delta \boldsymbol{r}$ as a reasonable approximation of the largest observable perturbation of the distance $\delta r$.

## 4. Discussion

### 4.1. Impulse Model

The results obtained from our simulations show that the perturbations induced by a halo of PBHs grow over time, to a median value of $|\delta \boldsymbol{r}|/r \sim 10^{-13}$ within 20 yr. To provide some analytical understanding of our results, we make use of the impulse approximation (J. Binney & S. Tremaine 2008) to compute the velocity change induced on an SSO by the gravitational pull from a PBH. It is assumed that the black hole moves in a straight line with an impact parameter (distance of closest approach) $b$. The motion of the SSO itself is neglected, as it is assumed that its orbital period is much longer than the timescale of the flyby. These assumptions are justified owing to the high velocities of the PBHs with $v_{\rm PBH} \sim 280$ km s$^{-1}$. The velocity change induced on an SSO in the direction perpendicular to the motion of the PBH is

$$\Delta v \approx \int_{-\infty}^{\infty} dt \frac{GM_{\rm PBH} b}{(b^2 + (v_{\rm PBH} t)^2)^{3/2}} = \frac{2GM_{\rm PBH}}{bv_{\rm PBH}}. \quad (4)$$

Note that in most cases that we discuss one has $v_{\rm PBH} t \gg b$, and thus the integration boundaries can safely be taken to infinity.

This is discussed in more detail in Section 4.3. In the following, we assume that all PBHs move perpendicular to the ecliptic plane, which is a first-order approximation that greatly simplifies our calculations and is also motivated by our initial conditions (see Section 2). This implies that the point of closest approach is identical for all bodies in the solar system and will be reached when the PBH crosses the ecliptic plane.

Let us first discuss the case where the impact parameter $b$ is much larger than the distance $r$ between Earth and another SSO (in our case Mars). Then, Earth and the other SSO will be perturbed by the same amount $\Delta v$ but in slightly different directions, with an angle between $\Delta \boldsymbol{v}_{\rm earth}$ and $\Delta \boldsymbol{v}_{\rm SSO}$ of $\alpha \approx r/b$. This means

$$|\delta \boldsymbol{r}| \approx |\Delta \boldsymbol{v}_{\rm earth} - \Delta \boldsymbol{v}_{\rm SSO}| t \approx \frac{\Delta v r t}{b} \approx \frac{2GM_{\rm PBH} r t}{b^2 v_{\rm PBH}}. \quad (5)$$

If, on the other hand, $b \ll r$, we only need to consider the velocity change of the body closest to the PBH,

$$|\delta \boldsymbol{r}| \approx |\Delta \boldsymbol{v}_{\rm earth} - \Delta \boldsymbol{v}_{\rm SSO}| t \approx \Delta v t = \frac{2GM_{\rm PBH} t}{bv_{\rm PBH}}. \quad (6)$$

In the intermediate case, the perturbation strength depends on the specific geometry. Numerically, by averaging over all possible geometries for a fixed $b$, we find that

$$\frac{|\delta \boldsymbol{r}|}{r} \approx \frac{2GM_{\rm PBH} t}{(br + b^2) v_{\rm PBH}} \quad (7)$$

provides a reasonable fit to the numerical data, with a maximum deviation of a factor of 2.2 at $b = r/2$. For the mean perturbation strength $\langle |\delta \boldsymbol{r}|/r \rangle$ that we discuss below the difference is at most 20%. The numerical result is discussed further in the Appendix.

So far, we have considered the perturbation from a single encounter with a PBH. For a DM halo composed of PBHs the rate of encounters in the solar system will be $\Gamma = (\sigma/m)\rho v = (\pi b^2 / M_{\rm PBH}) \rho_{\rm CDM} v_{\rm PBH}$. This means that the number of scattering events per interval of $b$ is $dN/db = (2\pi b / M_{\rm PBH}) \rho_{\rm CDM} v_{\rm PBH} t$.

From this we can obtain the perturbation strength,

$$\left\langle \frac{|\delta \mathbf{r}|}{r} \right\rangle \approx \int_0^{b_{\max}} db \, \frac{dN}{db} \frac{2GM_{\mathrm{PBH}}t}{(br+b^2)v_{\mathrm{PBH}}}$$
$$= 4\pi G \rho_{\mathrm{CDM}} t^2 \log(1 + b_{\max}/r). \quad (8)$$

Before we discuss this result in more detail, we have to specify the upper integration bound $b_{\max}$. In principle, the impact parameter $b$ can be as large as the box size $R_{\mathrm{box}}$ of our simulation. However, with increasing distance the number of PBH encounters grows until they will no longer cause individual perturbations but rather behave as a spherically symmetric mass distribution. Let us define $b_1$ as the impact parameter for which we expect one scattering event within the time $t$:

$$b_1 = \sqrt{\frac{M_{\mathrm{PBH}}}{\pi \rho_{\mathrm{CDM}} v_{\mathrm{PBH}} t}}. \quad (9)$$

Accordingly, there are $N^2$ encounters with an impact parameter $b \leqslant Nb_1$ within the same time period. However, as $N$ grows, these perturbations will begin to cancel each other out owing to the random orientation of $\delta v$. As can be seen in Figures 1 and 4 (left panel), the choice $b_{\max} = 3b_1$ provides a good fit to the actual mean perturbation, which suggests that the perturbation strength is well determined by the influence of the closest $N \sim \mathcal{O}(10)$ PBHs. Notably, if we set $b_{\max} = R_{\mathrm{box}}$, Equation (8) predicts a mean perturbation strength that is up to an order of magnitude larger than the simulation results and does not recover the correct dependence on $M_{\mathrm{PBH}}$. This also demonstrates that our choice of $R_{\mathrm{box}}$ has no significant effect on the outcome of the simulation.

The fact that the perturbations are dominated by the closest encounter is also demonstrated in the right panel of Figure 2, where we plot the perturbation strength $|\delta \mathbf{r}|/r$ at $t=1$ yr against the smallest impact parameter within the simulation run. The observed values agree reasonably well with Equation (7), which is displayed as dashed lines. Note that here we have assumed a singular encounter with a PBH that moves perpendicular to the ecliptic plane. Therefore, some deviation from the analytical estimates is expected in the more complex simulation environment. In addition, it is expected that other close encounters will also contribute to the perturbation strength. Note that the approximation of infinite integration boundaries in Equation (4) breaks down for large PBH masses, which explains part of the offset observed for $M_{\mathrm{PBH}} = 10^{21}$ g. This will be discussed in more detail in Section 4.3.

Importantly, Equation (8) provides a simple explanation for the observed weak dependence of the perturbation strength $|\delta \mathbf{r}|/r$ on the black hole mass, as it predicts only a logarithmic dependence as long as $b_{\max} \gg r$. The underlying reason is that the strength of the gravitational force ($\sim M_{\mathrm{PBH}}$) is counteracted by the rate of encounters ($\sim 1/M_{\mathrm{PBH}}$). However, if PBHs are sufficiently light (or $t$ sufficiently long), then we are in the limit $b_{\max} \ll r$ (assuming again that $b_{\max} \sim \mathcal{O}(b_1)$ as we argued above) and thus $|\delta \mathbf{r}|/r \sim \log(1+b_{\max}/r) \approx b_{\max}/r \sim \sqrt{M_{\mathrm{PBH}}}$. Indeed, the left panel of Figure 2 shows that the mean perturbation strength depends more strongly on $M_{\mathrm{PBH}}$ in the limit $b_{\max} \ll r$, in good agreement with our analytical model that is also displayed in the figure. The power-law slope appears to be more shallow than expected, but simulations with $M_{\mathrm{PBH}} < 10^{18}$ g are required to investigate this in more detail. Equation (8) also explains the lower values of $|\delta \mathbf{r}|/r$ for the outer planets in the right panel of Figure 4, as we have $b_{\max} < r$ and thus $|\delta \mathbf{r}|/r \sim 1/r$ there. Finally, we want to emphasize that the decreasing strength of the perturbations at low PBH masses provides consistency, as they should disappear as $M_{\mathrm{PBH}} \to 0$ ($\rho_{\mathrm{CDM}} = \rho_{\mathrm{PBH}} = \mathrm{const.}$) and a smooth DM component is obtained.

### 4.2. Statistics of the Perturbations

The impulse model derived in the previous section allows us to interpret the results obtained for the probability distribution $p(|\delta \mathbf{r}|/r)$. The homogeneous distribution of PBHs implies that the distribution of impact parameters is given by $p(b) \sim b$. In the case where $b \gg r$, Equation (5) can be used to obtain

$$p(|\delta \mathbf{r}|/r) = p(b) \left| \frac{db}{d|\delta \mathbf{r}|/r} \right| \sim (|\delta \mathbf{r}|/r)^{-2}. \quad (10)$$

In contrast, for $b \ll r$ it follows from Equation (6) that

$$p(|\delta \mathbf{r}|/r) = p(b) \left| \frac{db}{d|\delta \mathbf{r}|/r} \right| \sim (|\delta \mathbf{r}|/r)^{-3}. \quad (11)$$

The predicted power-law slopes of $-2$ and $-3$ in the respective limits of $b<r$ and $b>r$ are consistent with the data shown in Figure 3.

The slope of the distribution has notable consequences for the robustness of the mean perturbation strength that we obtain from the simulation data. In the case of Equation (10), the mean perturbation strength is logarithmically divergent, whereas it quickly converges for the case of Equation (11). Indeed, for simulations with $b_1 < r$ we find that the estimated uncertainty of the mean $\sigma/\sqrt{N}$, where $N$ is the number of simulation runs, is very low. In principle, by running enough simulations, the limit $b<r$ can be properly sampled for any value of $M_{\mathrm{PBH}}$. However, for large PBH masses this becomes computationally very expensive for the Earth–Mars distance. Nevertheless, we can estimate the uncertainty in the mean perturbation strength by computing the probability distribution $p(|\delta \mathbf{r}|/r)$ from Equation (7) using $p(b) \sim b$ and fitting it to the tail of the distributions obtained from simulation data. We can then compute the mean of a combined distribution, where we take the simulation data for the range of $|\delta \mathbf{r}|/r$ that are well sampled and the modeled tail for larger perturbations. The mean perturbation strength that we obtain in this way only differs by at most a factor of two from that obtained using only simulation data.

### 4.3. Extrapolating Our Results

Our simulations help us to investigate perturbations induced by PBHs for a certain range of masses $M_{\mathrm{PBH}}$ and timescales $t$. We find that the dependence on the mass of the PBH is logarithmic as long as $b_{\max} \sim b_1 \gg r$ and that, for the parameters that we studied, the results are well described by the impulse model, discussed in Section 4.1. Therefore, we can use our analytical model to extend the study to longer timescales and to a wider range of PBH masses that we could not investigate so far owing to the computational expense of running a large number of additional simulations. However, it is necessary to make some modifications to the impulse model. The reason is that for large PBH masses or on small timescales the PBHs will not act as a stream of particles but rather as a cluster with negligible movement. More precisely, the integration boundaries in Equation (4) cannot be taken to infinity if $v_{\rm PBH}t \gg b$ is not valid.

Taking into account the finite travel distance of the PBH, the velocity change induced on an SSO is given by

$$\Delta v(t) \approx \int_{-t/2}^{t/2} dt' \frac{GM_{\rm PBH}b}{(b^2 + (v_{\rm PBH}t')^2)^{3/2}}$$
$$= \frac{2GM_{\rm PBH}}{bv_{\rm PBH}} \frac{\frac{v_{\rm PBH}t}{2b}}{\sqrt{1 + \left(\frac{v_{\rm PBH}t}{2b}\right)^2}}, \quad (12)$$

which reduces to Equation (4) for $v_{\rm PBH}t \gg b$. The displacement of an SSO $\Delta r(t)$ can be computed by integrating Equation (12) with respect to time:

$$\Delta r(t) \approx \int_0^t dt' \frac{2GM_{\rm PBH}}{bv_{\rm PBH}} \frac{\frac{v_{\rm PBH}t'}{2b}}{\sqrt{1 + \left(\frac{v_{\rm PBH}t'}{2b}\right)^2}}$$
$$= \frac{4GM_{\rm PBH}}{v_{\rm PBH}^2}\left(\sqrt{1 + \left(\frac{v_{\rm PBH}t}{2b}\right)^2} - 1\right). \quad (13)$$

To obtain an explicit expression for the perturbation strength, we must take into account that for $b \gg r$ one has $|\delta r| \approx r/b\, \Delta r$, as we discussed in Section 4.1. Equivalently to Equation (7), a general expression can be used to good approximation. It then follows that the perturbation strength from a single PBH is given by

$$\frac{|\delta r|}{r} \approx \frac{4GM_{\rm PBH}}{(r+b)v_{\rm PBH}^2}\left(\sqrt{1 + \left(\frac{v_{\rm PBH}t}{2b}\right)^2} - 1\right). \quad (14)$$

In the limit $v_{\rm PBH}t \gg b$ this expression reduces to Equation (7), as expected. For $v_{\rm PBH}t \ll b$ we have

$$\frac{|\delta r|}{r} \approx \frac{GM_{\rm PBH}t^2}{2(rb^2 + b^3)}. \quad (15)$$

To demonstrate the effect of this modification, we show this improved model as a dotted line for $M_{\rm PBH} = 10^{21}$ g in the right panel of Figure 2. Taking into account the finite travel distance leads to better agreement with the simulation data and is important in the limit $v_{\rm PBH}t \ll b$.

For $v_{\rm PBH}t \gg b$ the PBHs behave like a stream of particles with $dN/db = (2\pi b/M_{\rm PBH})\rho_{\rm CDM}v_{\rm PBH}t$. On the other hand, if $v_{\rm PBH}t \ll b$, the PBHs behave as a cluster with negligible movement and the distribution of their distances is given by $dN/db = 4\pi b^2 \rho_{\rm CDM}/M_{\rm PBH}$. By studying the numerical distribution $dN/db$ from our simulation data, we found that

$$\frac{dN}{db} \approx \frac{2\pi b \rho_{\rm CDM}}{M_{\rm PBH}}(v_{\rm PBH}t + 2b) \quad (16)$$

provides a good approximation in the general case. Using this, the mean perturbation strength is given by

$$\left\langle \frac{|\delta r|}{r} \right\rangle = \int_0^{b_{\rm max}} db \, \frac{dN}{db} \frac{|\delta \mathbf{r}|}{r}$$
$$\approx \int_0^{b_{\rm max}} db \, \frac{8\pi \rho_{\rm CDM} Gb(v_{\rm PBH}t + 2b)}{(r+b)v_{\rm PBH}^2}$$
$$\left(\sqrt{1 + \left(\frac{v_{\rm PBH}t}{2b}\right)^2} - 1\right). \quad (17)$$

To include only the closest $\mathcal{O}(10)$ encounters, we set $b_{\rm max} = 3b_1$ with $b_1$ now given by

$$b_1 = \min\left\{\left(\frac{M_{\rm PBH}}{\pi \rho_{\rm CDM} v_{\rm PBH}t}\right)^{1/2}, \left(\frac{3M_{\rm PBH}}{4\pi \rho_{\rm CDM}}\right)^{1/3}\right\}. \quad (18)$$

We display the mean perturbation strength that follows from Equation (17) in the left panel of Figure 2. In the respective mass range, the difference to Equation (8) is insignificant and below the accuracy of the model and the simulation results. This is also the case for Figure 1 and the left panel of Figure 4, where the analytical model is shown as well. Importantly, for $v_{\rm PBH}t \ll b$, Equation (17) converges to $|\delta r|/r = 2\pi G\rho_{\rm CDM}t^2 \log(1 + b_{\rm max}/r)$, a factor of two smaller than Equation (8). This demonstrates that Equation (8) provides a reasonable estimate of the mean perturbation strength even in the limit $v_{\rm PBH}t \ll b$.

### 4.4. Prospects for Detecting PBHs

The quantity of interest for detecting PBHs is the ratio between the perturbations $|\delta r|$ that they induce and the observational residual $\sigma_r$ in the distance between Earth and a given SSO. In Section 4.2 we found that the distribution of perturbations $p(|\delta r|/r)$ is highly skewed. This implies that the mean perturbation strength is not a good measure of detectability, as the likelihood to observe it can be low. Of the 500 simulations that have been performed for $t = 20$ yr, 112 reach a perturbation strength $|\delta r|/r$ that is larger than the mean $\langle|\delta r|/r\rangle$, a fraction of $\sim 27\%$. Therefore, it is more useful to compare the median perturbation strength to the observational residual. If $|\delta r|_{\rm median} > \alpha \sigma_r$, then there is 50% chance for an observation with a signal-to-noise ratio (SNR) of $\alpha$. In Sections 4.1 and 4.3 we have derived the impulse model to describe the mean perturbation strength. To a good approximation, the median perturbation strength can be estimated by setting the minimum impact parameter to $b_1$ Equation (9) when performing the integration over $b$,

$$\left(\frac{|\delta r|}{r}\right)_{\rm median} \approx \int_{b_1}^{b_{\rm max}} db \, \frac{8\pi \rho_{\rm CDM} Gb(v_{\rm PBH}t + 2b)}{(r+b)v_{\rm PBH}^2}$$
$$\times \left(\sqrt{1 + \left(\frac{v_{\rm PBH}t}{2b}\right)^2} - 1\right). \quad (19)$$

The reason is that the median perturbation is dominated by the smallest "typical" impact parameter $b_1$, whereas the mean takes into account smaller values of $b$ that can occur rarely and is thus skewed to larger values. We demonstrate the validity of this assumption in the left panel of Figure 4, where we show the result from Equation (19) as the green line, which agrees well with the median from the simulation data (black dashed line).

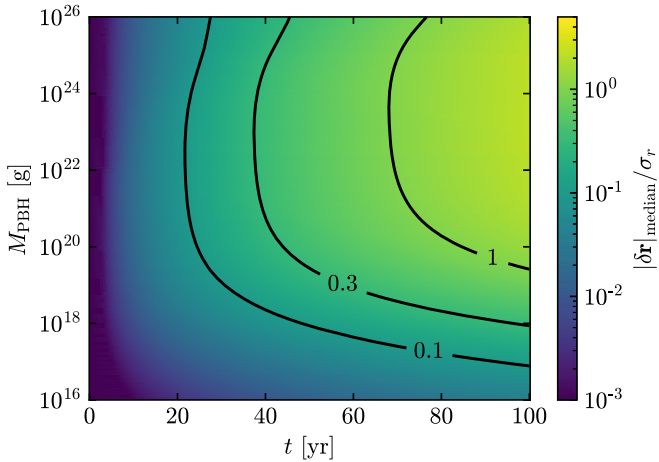

**Figure 5.** Estimated median SNR $|\delta r|_{\text{median}}/\sigma_r$ for the observation of the Earth–Mars distance as a function of the PBH mass and the timescale, assuming $\sigma_r = 70$ cm. Contour levels are drawn for several values of the SNR. The results are obtained using the impulse model, presented in Section 4.1 and refined in Section 4.3.

Figure 5 presents the SNR $|\delta r|_{\text{median}}/\sigma_r$ for the Earth–Mars distance using Equation (8) and $b_{\max} = 3b_1$, although the results are not very sensitive to the precise value of $b_{\max}$ as discussed in Section 4.1. A value of $\sigma_r = 70$ cm is assumed for the residual of the Earth–Mars distance (R. S. Park et al. 2021). Note that the dependence of $|\delta r|_{\text{median}}$ on $M_{\text{PBH}}$ is weak for $M_{\text{PBH}} \gtrsim 10^{20}$ g and the timescales considered here. At lower masses, when $b_{\max} \ll r$, one has $|\delta r|_{\text{median}} \sim \sqrt{M_{\text{PBH}}}$ and the perturbation strength decreases. In the limit $r \ll vt \ll b$ Equation (19) becomes independent of the black hole mass and converges to $|\delta r|_{\text{median}}/r = 2\pi G \rho_{\text{CDM}} t^2 \log(b_{\max}/b_1) \approx 3 \times 10^{-16}(t/1\,\text{yr})^2$. For $t = 20$ yr this limit is reached for $M_{\text{PBH}} \gg 10^{25}$ g, and for $t = 1$ yr the median perturbation strength becomes constant for $M_{\text{PBH}} \gg 10^{21}$ g. To test this, we have performed 9000 simulation runs for $M_{\text{PBH}} \in [10^{24}, 10^{25}]$ g with $R_{\text{box}} \in [4500, 10{,}000]$ au. The median perturbation strength that we obtain in both simulations is almost identical and agrees within a factor of two with the prediction from the model for all planets.

As a conservative estimate, we regard the perturbations induced by the halo of PBHs as a random signal, and thus an SNR greater than 1 is required for detection. We find that a significant detection ($|\delta r|_{\text{median}}/\sigma_r \geqslant 3$) cannot be reached with 100 yr of observational data. If the uncertainty in the distance $\sigma_r$ decreases by a factor of 30, then only around 20–30 yr of data are required to detect PBHs with a mass $M_{\text{PBH}} > 10^{19}$ g with $3\sigma$ confidence (this corresponds to the 0.1 contour in Figure 5). Note that $|\delta r| \sim \rho_{\text{CDM}}$ and hence the detection window becomes larger if the local DM density is greater than the assumed value of $\rho_{\text{CDM}} = 7 \times 10^{-25}$ g cm$^{-3}$. We want to emphasize that the results presented in Figure 5 are based on the analytical impulse model and for particular parameters one should perform a set of numerical simulations to obtain more accurate results. It would also be interesting to study the probability of detection for a given SNR as a function of $M_{\text{PBH}}$ and $t$. However, this requires a better understanding of $p(|\delta r|/r)$ that is beyond the scope of this work.

So far, we have treated the perturbations as a random signal and thus required an SNR > 1 for detection. However, the perturbations induced by the encounters of PBHs likely have certain characteristics that can be exploited to make predictions of the expected signal $\delta r(t)$. In that case, one can attempt to use template matching techniques in order to detect perturbations with an amplitude well below the noise level $\sigma_r$. This was studied by T. X. Tran et al. (2024), who found that there will be a sizable number of detectable encounters if the extraction of signals with an SNR well below $10^{-2}$ is achieved. A possible caveat regarding the detection of PBHs are potential degeneracies between the perturbations induced by the PBHs and other gravitational effects in the solar system. Accurate models of solar system ephemerides are obtained by iterating over a large number of free parameters, including the physical parameters of all SSOs. To assess the prospects of detecting PBHs, one has to study the degree of degeneracy between the perturbations that they induce and a change in other parameters of these models.

### 4.5. Poissonian Fluctuations

In our work, we study the cumulative gravitational effect of PBHs on the orbits of SSOs. We find that the perturbations induced by the black holes are dominated by the closest encounter (see Section 4.1). This is in contrast to the work by A. Loeb (2024), which studies the influence of Poissonian fluctuations in the PBH density. If the DM is composed of asteroid-mass black holes, then their number within the solar system can undergo sizable Poisson fluctuations. A. Loeb (2024) considered the rate of change in the total mass of PBHs within a distance $R$ from the Sun, given by

$$\delta \dot{M} = 1.9 \times 10^{-13} \left( \frac{M_{\text{PBH}}}{10^{20}\,\text{g}} \right)^{1/2} \left( \frac{R}{50\,\text{au}} \right)^{1/2} M_\odot\,\text{yr}^{-1}. \quad (20)$$

In his work it is then assumed that such fluctuations would induce similar effects to those if a point mass in the center of the solar system were to change its mass with the same rate, effectively changing the solar mass by a small amount. For $R = 50$ au and $M_{\text{PBH}} \in [6 \times 10^{18}, 10^{22}]$ g, this would lead to a rate $\delta \dot{M}$ larger than the constraint on $\dot{M}_{\text{sol}}$ by E. V. Pitjeva et al. (2021), and thus he concluded that PBHs in this mass range cannot make up the entire DM.

It is important to note that these Poissonian fluctuations can only occur if there is at least one PBH in the volume of interest. For Mars this requires $M_{\text{PBH}} \lesssim 3 \times 10^{16}$ g to have at least one PBH inside its orbit. At these low PBH masses the fluctuations that follow from Equation (20) are well below the observational constraint from E. V. Pitjeva et al. (2021) and thus cannot be used to rule out PBHs as DM. For the outer planets, the fluctuations from Equation (20) can exceed the constraint on $\dot{M}_{\text{sol}}$, as, e.g., for Neptune this requires $M \lesssim 3 \times 10^{20}$ g, which implies a value of $\delta \dot{M} < 2.5 \times 10^{-13}\,M_\odot\,\text{yr}^{-1}$. However, it is not justified to apply the constraint by E. V. Pitjeva et al. (2021) for the outer planets in isolation, as it was derived from the entire set of solar system data, assuming no dependence of $\dot{M}_{\text{sol}}$ on $r$, and is dominated by the effect of fluctuations on the inner planets. On the one hand, this is because the observational data of the inner planets currently have the highest accuracy. On the other hand, a change in the solar mass has a much stronger effect on the inner planets compared to the outer ones, as the additional force from the change in the solar mass will be proportional to $\delta \dot{M}/r^2$. In essence, an observational constraint on $\dot{M}_{\text{sol}}$ at the orbital distance of Neptune would be orders of magnitude weaker than the limit that A. Loeb (2024) is using. Note that similar remarks were already made by J. M. Cline (2024).

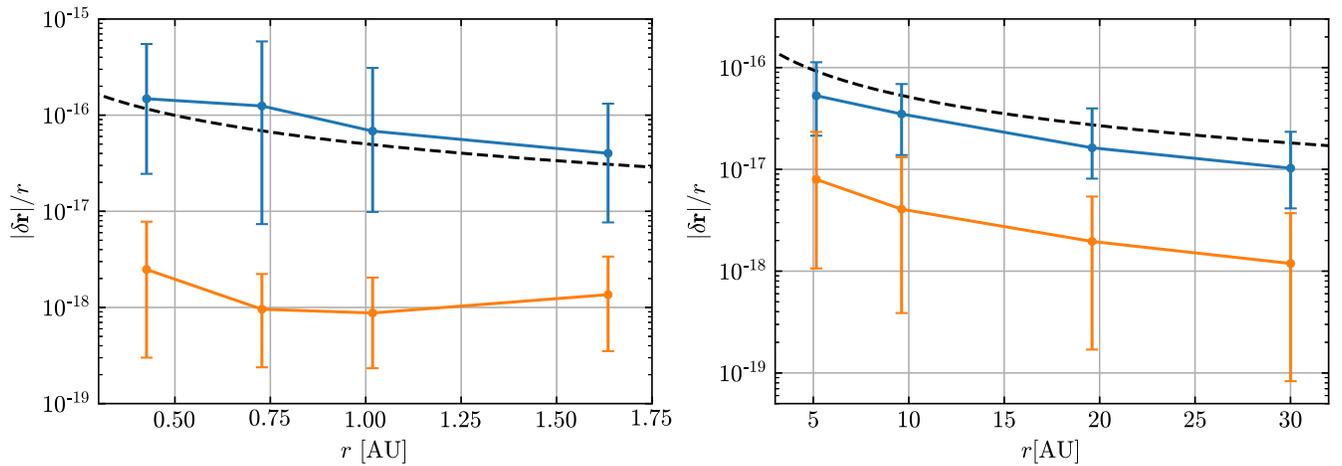

**Figure 6.** Perturbation of the vector between the Sun and the planets in the solar system. The left panel shows the results for a simulation of the inner solar system with $M_{\rm PBH} = 3 \times 10^{13}$ g at $t = 3$ yr. In the right panel, we present the results for the outer solar system with $M_{\rm PBH} = 10^{17}$ g at $t = 1$ yr. Each data point corresponds to the median perturbation of 100 simulation runs. The error bars indicate the range that includes 90% of the simulation results. The blue points correspond to our usual simulation setup. In orange, we show the results of simulations where the solar mass is changing according to the Poisson fluctuations of PBHs $\delta \dot{M}$, as explained in Section 4.5. The black dashed line is the analytical prediction of the median perturbation strength from the impulse model, according to Equation (19).

For the reasons stated above, one cannot rule out PBHs based on the Poisson fluctuation in their number density. Nevertheless, it remains of interest to compare the effects of these fluctuations to those from individual encounters. To investigate this, we perform a simulation where, at each time step and for each planet, we count the number of PBHs $N_{\rm PBH}(t, r_i)$ within the orbital distance $r_i$ of the planet $i$ to the Sun. We then set the solar mass for each planet to $M_{\rm sol,i}(t) = M_\odot + M_{\rm PBH} N_{\rm PBH}(t, r_i)$. The PBHs do not exert any gravitational force on the solar system bodies in this simulation, acting only through the fluctuations of the solar mass. There needs to be at least one PBH inside the orbit of a planet to observe any fluctuations. For Mercury this implies $M_{\rm PBH} \lesssim 4 \times 10^{14}$ g, which would lead to a computationally infeasible number of PBHs within the entire solar system. We therefore study the inner and outer planets in two separate simulations. For the simulation containing the inner planets and Earth's Moon we choose $M_{\rm PBH} = 3 \times 10^{13}$ g and $L_{\rm box} = 10$ au, whereas we take $M_{\rm PBH} = 10^{17}$ g and $L_{\rm box} = 150$ au for the outer planets. This means that there will be on average $\langle N_{\rm PBH} \rangle \in [13, 120, 347, 858]$ PBHs inside the orbits of Mercury, Venus, Earth, and Mars. For Jupiter, Saturn, Uranus, and Neptune there will be $\langle N_{\rm PBH} \rangle \in [12, 89, 737, 2622]$ PBHs inside their orbits. In Figure 6 we display the perturbations of the vector between the Sun and each planet. The left panel shows the results for the inner solar system at $t = 3$ yr, whereas the simulations for the outer planets are shown in the right panel for $t = 1$ yr. Blue points indicate the median value for simulations where we include the gravity of the PBHs, whereas the orange points show the results for the simulations with a fluctuating solar mass. The perturbations induced by PBHs on the orbit of the planets are roughly an order of magnitude larger compared to the effect of Poisson fluctuations of the solar mass. Based on these results alone, we cannot distinguish whether the disagreement is because the Poisson fluctuations are a small contribution to the total effect of PBHs in the solar system or because one cannot treat the Poisson fluctuations of the PBH number density as a change in the solar mass. However, the analytical model introduced in Section 4.1 suggests that the perturbations induced by PBHs can be predicted with good accuracy by taking into account the closest encounters. We display the model for the median perturbation strength from Equation (19) as a dashed line in Figure 6, and it shows good agreement for $b_{\rm max} = 3b_1$. This suggests that Poisson fluctuations likely play only a minor role in the overall gravitational effect of PBHs on the solar system bodies.

### 4.6. Limitations of This Work

Our simulations enable us to make general statements about the strength and evolution of perturbations induced by PBHs and to study whether they exceed current detection limits. However, in order to actually detect individual PBHs, one needs a precise model for solar system ephemerides that includes more solar system bodies and treatment of finite-size effects, radiation pressure, relativistic corrections, and other effects that we have neglected. It is crucial to study the possible degeneracy between the mentioned effects and the perturbations induced by the PBHs. A combination of such a model with a sophisticated analysis of solar system *data* could be a viable pathway to the detection of asteroid-mass PBHs.

For the halo of PBHs, we have assumed a monochromatic mass function for simplicity. In reality, PBHs are likely to form with an extended mass distribution. While one can easily expand our analysis for a distribution of PBH masses, it is not straightforward to connect this distribution to the initial PBH mass function at formation in the early Universe. The main reason for this is that one has to deal with migration effects, where the heavier black holes sink toward the center of the halo while the lighter ones move outward. In general, we do not expect that an extended mass function would change the perturbations of PBHs in a drastic way, as we found little mass dependence in our results.

A more sizable impact on our results could be expected if PBHs are clustered (see Section II of B. J. Carr et al. 2024 and references therein). If the size of such a cluster is small compared to the impact parameter, then it will simply act as a much heavier black hole. However, if the size of such a cluster is comparable to the solar system, then our results could change in a nontrivial way. In addition, the degree of clustering might be dependent on the black hole mass, in which case an extended mass function could have a sizable impact on the result. Therefore, it would be worthwhile to study this aspect in a future work.

## 5. Summary

In this work, we perform numerical simulations of the solar system, embedded in a DM halo of PBHs. The simulation results are used to quantify the perturbations induced by the PBHs on the distance between Earth and another SSO. First, it is demonstrated that the strength of the perturbations depends only logarithmically on the PBH mass if the nearest PBH is much farther away from Earth than the other SSO. The physical reason for this is the fact that the strength of the gravitational force ($\sim M_{\rm PBH}$) is balanced by the rate of encounters ($\sim 1/M_{\rm PBH}$). We then show for $M_{\rm PBH} = 10^{20}$ g that after a time span of 20 yr the perturbations are still more than an order of magnitude below the current precision of observational data for the Earth–Mars distance, which has been available for a similar time period. Therefore, PBHs cannot be directly constrained based on solar system ephemerides, challenging the results of recent work.

Our findings are interpreted by a simple analytical model that provides an independent test of the accuracy of our simulations. It allows us to show that the perturbations induced by the PBHs are dominated by the closest encounter. We find that the effect of Poissonian fluctuations in the PBH number density studied by A. Loeb (2024) is significantly smaller. In addition, the analytical model enables us to extrapolate our results to longer timescales and to a larger range of PBH masses. We find that the precision of solar system ephemerides has to increase by roughly an order of magnitude for the effect of the PBH halo to become noticeable within a decade of observation. Equivalently, PBHs could be detected using already-existing data if the extraction of signals well below the noise level is achieved. Importantly, in order to reliably detect PBHs, it is necessary to employ accurate models of solar system ephemerides that include finite-size effects, relativistic corrections, and other effects that were neglected in this work.

## Acknowledgments

We thank Avi Loeb, whose work has inspired us to pursue this project. We also would like to thank Chris McKee and Florian Kühnel for fruitful discussions and helpful comments. This research was supported by the Excellence Cluster ORIGINS, which is funded by the Deutsche Forschungsgemeinschaft (DFG, German Research Foundation) under Germany's Excellence Strategy – EXC-2094 – 390783311.

*Software:* Julia (J. Bezanson et al. 2017), Matplotlib (J. D. Hunter 2007).

## Appendix
## Relation between the Perturbation Strength and the Impact Parameter

In Section 4.1 we derived expressions for the relation between the perturbation strength $|\delta \mathbf{r}|/r$ and the impact parameter $b$ in the limit $b \gg r$ Equation (5) and for $b \ll r$ (Equation 6). In the intermediate case $b \sim r$ the result will strongly depend on the precise geometry of the encounter. To investigate this effect, we uniformly sample $10^7$ PBH positions in a plane around two SSOs that are separated by distance $r = 1$. For each PBH position we determine the minimum of the distance to both bodies $b$ and the perturbation strength $|\delta \mathbf{r}|/r$. Figure 7 shows the range of values obtained for $|\delta \mathbf{r}|$ for each value of $b$ as the shaded area and the mean values as points. We have set $2GM_{\rm PBH}\Delta t/v_{\rm PBH} = r = 1$, as we only care about the functional dependence on $b$. The analytical relation (Equation (7)) that we use in this work is shown as the orange line and agrees well with the numerical data. The strongest deviation is a factor of 2.2 for $b \approx r/2$. If we integrate over the numerical relation, then we find a deviation of at most 20%. This justifies using Equation (7) to describe the simulation results.

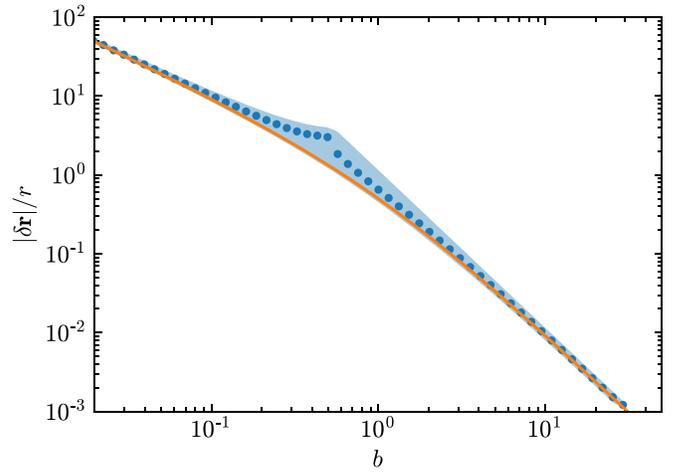

**Figure 7.** Strength of the perturbation $|\delta \mathbf{r}|/r$ induced by a flyby of a PBH as a function of the impact parameter $b$ in units of $r$. The shaded area displays the range of values for all possible geometries. The blue circles indicate the mean value averaged over all configurations. The analytical relation from Equation (7) is shown as the orange line.

## ORCID iDs

Valentin Thoss ⬤ https://orcid.org/0009-0007-5515-2158
Andreas Burkert ⬤ https://orcid.org/0000-0001-6879-9822